\documentclass[epj]{svjour}
\usepackage{amsmath,amssymb} \usepackage{graphicx}

\begin{document}
\title{Force distributions and force chains in random stiff fiber networks}
\author{Claus Heussinger\inst{1} \and Erwin Frey\inst{1}}
\institute{Arnold Sommerfeld Center for Theoretical Physics and Center for
  NanoScience (CeNS), Department of Physics, Ludwig-Maximilians-Universit\"at
  M\"unchen, Theresienstrasse 37, D-80333 M\"unchen, Germany}
\date{\today}
\abstract{
  We study the elasticity of random stiff fiber networks. The elastic response
  of the fibers is characterized by a central force stretching stiffness as well
  as a bending stiffness that acts transverse to the fiber contour. Previous
  studies have shown that this model displays an anomalous elastic regime where
  the stretching mode is fully frozen out and the elastic energy is
  completely dominated by the bending mode. We demonstrate by simulations and
  scaling arguments that, in contrast to the bending dominated \emph{elastic
    energy}, the equally important \emph{elastic forces} are to a large extent
  stretching dominated. By characterizing these forces on microscopic,
  mesoscopic and macroscopic scales we find two mechanisms of how forces are
  transmitted in the network. While forces smaller than a threshold $F_c$ are
  effectively balanced by a homogeneous background medium, forces larger than
  $F_c$ are found to be heterogeneously distributed throughout the sample,
  giving rise to highly localized force-chains known from granular media.
\PACS{ {62.25.+g}{Mechanical properties of nanoscale materials} \and
  {87.16.Ka}{Filaments, microtubules, their networks, and supramolecular
    assemblies} %\and {81.05.Lg}{Polymers and plastics; rubber; synthetic and
    %natural fibers; organometallic and organic materials} 
} % end of PACS codes
} %end of abstract
\maketitle
\section{Introduction}
\label{intro}

\begin{figure}[t]
   \begin{center}
  \includegraphics[width=0.85\columnwidth]{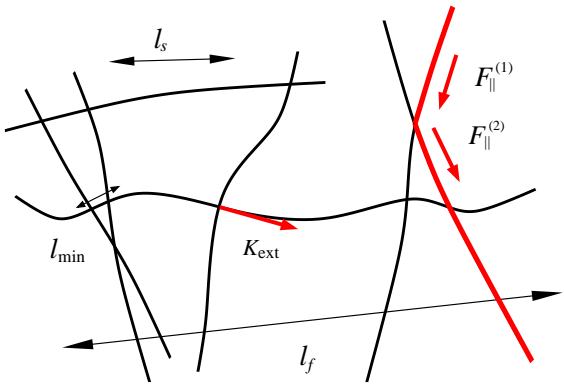}
   \end{center}
   \caption{Illustration of the local network structure and the relevant
     length-scales in the random fiber network (drawn in the deformed
     configuration): the fiber length $l_f$, the typical segment length $l_s$
     and the shortest deformed segment of length $l_{\rm min}$. $K_{\rm ext}$
     signifies the bending force that the crossing fiber exerts in the axial
     direction of the horizontal fiber.  $F_\parallel^{(1)}$ and
     $F_\parallel^{(2)}$ correspond to axial forces that are directly
     transmitted from one fiber to a neighboring fiber at the crosslink. This
     mechanism forms the basis for the establishment of force-chains; see main
     text.}
   \label{fig:illustration}
\end{figure}

It has been well known for more than a century that networks of central force
springs lose their rigidity when the number of connected neighbours is lower
than a certain threshold value~\cite{maxwell1864}.  To guarantee the rigidity of
these otherwise ``floppy'' networks, additional contributions to the elastic
energy, beyond central-force stretching, have to be introduced~\cite {sahimi}.
Here our focus is on a particular class of heterogeneous networks composed of
crosslinked fibers, whose length $l_f$ is much larger than the typical distance
$l_s$ between two fiber-fiber intersections (see Fig.~\ref{fig:illustration}).
These systems have recently been suggested as model systems for studying the
mechanical properties of paper sheets~\cite{alava06} or biological networks of
semiflexible polymers~\cite{frey98,bausch06,heu06a}.  As only two fibers may
intersect at a given cross-link, the average number of neighboring cross-links
is $z<4$. This places them below the rigidity transition both in two and in
three spatial dimensions.  Several strategies have been used to elastically
stabilize a central-force fiber network~\cite{lat01springs}. Here, we use an
additional energy cost for fiber ``bending''. The bending mode penalizes
deformations transverse to the contour of the fiber while to linear order the
distance between cross-links, i.e. the length of the fiber, remains unchanged.

The two-dimensional fiber network we consider is defined by randomly placing $N$
initially straight elastic fibers of length $l_f$ on a plane of area $A=L^2$
such that both position and orientation are uniformly distributed. The fiber
density is thus defined as $\rho = Nl_f/A$. We consider the fiber-fiber
intersections to be perfectly rigid, but freely rotatable cross-links that do
not allow for relative sliding of the fibers. The elastic building blocks of the
network are the fiber segments, which connect two neighbouring cross-links. A
segment of length $l_s$ is modeled as a classical beam with cross-section radius
$r$ and bending rigidity $\kappa$~\cite{landau7}.  Loaded along its axis
(central force ``stretching'') such a slender rod has a rather high stiffness,
characterized by the spring constant $k_\parallel(l_s) \sim \kappa/l_sr^2$,
while it is much softer with respect to transverse deformations
$k_\perp(l_s)\sim \kappa/l_s^3$ (``bending'').

While this random fiber network is known to have a rigidity percolation
transition at a density $\rho_c$~\cite{lat01,wil03,hea03b}, we have recently
shown~\cite{heu06floppy,heu07condmat} how the network's inherent fragility,
induced by its low connectivity $z$, determines the properties even in the
high-density regime far away from the percolation threshold, $\rho \gg \rho_c$.
In particular, it was possible to explain the anomalous scaling properties of
the shear modulus as found in simulations~\cite{wil03,hea03a}. The unusually
strong density dependence of the elastic shear modulus $G\sim
(\rho-\rho_c)^{6.67}$~\cite{wil03} is found to be a consequence of the
architecture of the network that features various different
length-scales~\cite{heu06floppy,heu07condmat} (see Fig.~\ref{fig:illustration}).
On the mesoscopic scale the fiber length $l_f$ induces a highly non-affine
deformation field, where segment deformations $\delta_{\rm na}$ follow the
macroscopic strain $\gamma$ as $\delta_{\rm na}\sim \gamma l_f$.  This is in
stark contrast to an affine deformation field where deformations scale with the
size of the object under consideration. For a segment of length $l_s$ the affine
deformation therefore is $\delta_{rm aff}\propto \gamma l_s$. Microscopically, a
second length $l_{\min}$ governs the coupling of a fiber segment to its
neighbours on the crossing fiber~\cite{heu06floppy,heu07condmat}.  Due to the
fact that the bending stiffness $k_\perp\sim l_s^{-3}$ of a segment is strongly
increasing with shortening its length, it is found that segments with $l_s<l_
{\rm min}$ rather deform their neighbours than being deformed, while longer
segments $l_s>l_{\rm min}$ are not stiff enough to induce deformations in the
surrounding. Thus $l_{\rm min}$ plays the role of a \emph{rigidity scale}, below
which segments are stiff enough to remain undeformed.  As a consequence the
elastic properties of the fiber as a whole are not governed by the
\emph{average} segment $\bar l_s$, but rather by the \emph{smallest loaded}
segment $l_{\min}$.

In the observed scaling regime the elastic modulus does originate exclusively in
the bending of the individual fibers, and thus reflects the stabilising effect
of this soft deformation mode. In contrast, stretching deformations in this
non-affine bending dominated regime can be assumed to be frozen out completely.
In our previous articles~\cite{heu06floppy,heu07condmat} we have dealt with the
properties of the bending mode, $k_\perp$, and its implications on the \emph
{elastic energy}.  Here we focus on the stretching mode, $k_\parallel$, and its
role in the occurence of \emph{elastic forces}. As the non-affine bending regime
is relevant for slender fibers with a small cross-section radius, $r\to 0$, it
is characterized by a scale separation, $k_\parallel/k_\perp\sim
r^{-2}\to\infty$, which assures that no stretching deformations,
$\delta_\parallel$, occur. Thus, the fibers effectively behave as if they were
inextensible bars.  Closer inspection of the limiting process reveals, however,
that the stretching deformations tend to zero as $\delta_\parallel \sim
1/k_\parallel$~\footnote{This may be seen by considering the following
  simplified energy function, $W = k_\perp(\delta-\delta_\parallel)^2 +
  k_\parallel\delta_\parallel^2 $, which represents a system of two springs
  connected in series.  Minimizing the energy for fixed overall deformation
  $\delta$, one finds $\delta_\parallel = k_\perp\delta/(k_\parallel+k_\perp)$,
  which shows the reqired scaling $\delta_\parallel \sim k_\parallel^{-1}$ in
  the limit $k_\parallel\to\infty$.}.  This makes the contribution to the total
energy $W_s \sim k_\parallel\delta_\parallel^2 \sim k_\parallel^{-1}$
negligible, as required, but also implies that finite stretching forces
$F_\parallel$ will occur,
\begin{equation}\label{eq:finiteStretchForce}
   F_\parallel\sim k_\parallel\delta_\parallel\to \rm const\,.
\end{equation}
Indeed, these forces, acting axially along the backbone of the fibers,
are absolutely necessary to satisfy force-balance at the intersection
of two fibers.  With two fibers intersecting at a finite angle it is
intuitively clear that a change in transverse force in one fiber has
to be balanced by an axial force in the second.

For thicker fibers with a larger cross-section radius $r$, a second elastic
regime occurs, where the bending instead of the stretching mode is frozen out.
This regime formally corresponds to the limit $k_\parallel/k_\perp \sim
r^{-2}\to0$. It is characterized by stretching deformations of mainly affine
character~\cite{hea03a}. The elastic shear modulus (as well as the Young's
modulus) have been shown to depend linearly on density~\cite{wil03,ast00,ast94},
$G\sim \rho$, which is in striking contrast to the strong susceptiblity to
density variations found in the non-affine bending regime.

In the following we will present results of simulations that characterize in
detail the occuring axial forces in both elastic regimes, the non- affine
bending as well as the affine stretching regime. In the simulations we subject
the network to a macroscopic deformation and determine the new equilibrium
configuration by minimizing the elastic energy. The minimization procedure is
performed with the commercially available finite element solver MSC.MARC.
Further details of the simulation procedure can be found in our previous
publications~\cite{heu06a}. Starting with the average force-profile along the
fibers we then proceed by giving the full probability distribution of forces.
We show that its tails are heterogeneously distributed throughout the system,
similar to force-chains in granular media. We find that most of the features can
be understood with the help of the two basic length-scales, the filament length
$l_f$ and the rigidity scale $l_{\min}$.

\section{Effective Medium Theory}\label{sec:cox-model}

In this section we will characterize the configurationally averaged
force profile along a fiber. In the spirit of an effective medium
theory, one can think of the fiber as being imbedded in an elastic
matrix that, on a coarse-grained scale, acts continuously along the
backbone. The associated force ${\cal K}_{\rm ext}$, which is imposed
on the fiber, leads to a change in axial force according
to~\cite{landau7}
\begin{equation}\label{eq:landauAxial} \frac{\partial F_\parallel} 
{\partial s} =
-{\cal K}_{\rm ext}\,,
\end{equation}
where $s$ is the arclength along the fiber backbone.

A while back, Cox~\cite{cox51} provided a second, constitutive equation that
allows to solve for the force profile $F_\parallel(s)$.  He assumed the medium
to be characterized by an affine deformation field $\delta_{\rm aff}(s) \sim
\gamma s$. The external force ${\cal K}_{\rm ext}$ is assumed to be non-
vanishing only when the actual fiber deformation $\delta_\parallel$ is different
from this affine deformation field,
\begin{equation}\label{eq:coxMedium}
   {\cal K}_{\rm ext}(s) 
   = k \left(\delta_\parallel-\delta_{\rm aff}(s)\right)\,.
\end{equation}

Eqs.~(\ref{eq:landauAxial}) and (\ref{eq:coxMedium}) can easily be solved and
result in a force profile that shows a plateau in the center of the fiber as
well as boundary layers where the force decreases exponentially, $
F_\parallel(s) = a-b\cosh[c(s-l_f/2)]$, with $a$, $b$, and $c$ appropriately
chosen constants.  

\begin{figure}[thb]
  \begin{center}
    \includegraphics[width=0.9\columnwidth]{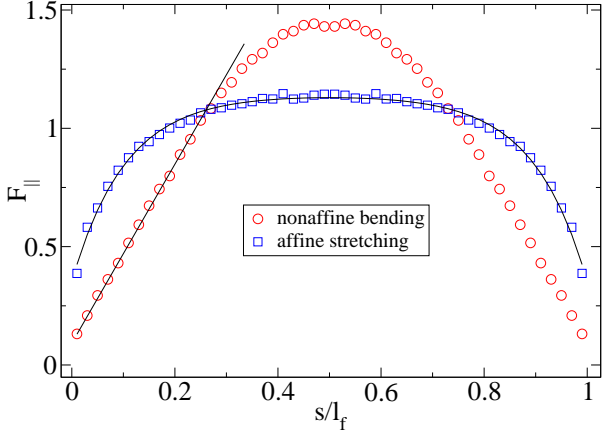}
\end{center}
\caption{Variation of average axial force $F_\parallel$ along the backbone
  $s=[0,l_f]$ of the fibers (symmetrized around $s=l_f/2$).  The symbols are the
  results of our simulations. Towards the fiber ends the force relaxes
  exponentially in the affine stretching regime (full curve is a fit to the Cox
  model), and linearly in the non-affine bending regime.}
   \label{fig:stressProfile}
\end{figure}

{\AA}str{\"o}m et al.~\cite{ast94} have applied this model to the affine
stretching regime and found the boundary layers to grow as the fiber
cross-section radius is decreased.  Fig.~\ref{fig:stressProfile} shows results
of our simulations for the force profile both in the affine stretching regime
(blue squares) and the non-affine bending regime (red circles).  Apparently the
Cox-model accounts very well for the force profile in the stretching regime,
while it fails completely in the bending regime, where the simulation data
clearly show that the force increases linearly from the boundary towards the
center of the fiber.

Cox ideas can be generalized to the non-affine bending regime, where the elastic
medium entirely consists of bending modes. In this regime the axial forces in
the fiber arise due to the pulling and pushing of its crossing neighbors that
try to transfer their high bending load in a kind of lever-arm effect (see
Fig.\ref {fig:illustration}). As explained above, the deformation field is
non-affine and, instead of $\delta_{\rm aff}$, one has to use $\delta_{\rm na}
\sim \gamma l_f$~\cite{heu07condmat}, which is proportional to the fiber length
$l_f$.  Since we are interested in the limit where stretching deformations
vanish, $\delta_\parallel\to 0$, the resulting exernal force is arclength
independent, ${\cal K}_{\rm ext} = -k\delta_{\rm na}\sim -l_f$ and constant
along the backbone. The axial force profile $F_\parallel(s)$ is thus expected to
be linearly increasing from the boundaries towards the center, in agreement with
the simulations.

Recently, Head et al.~\cite{hea03a} have suggested growing boundary layers to
play a key role for the cross-over from the affine stretching to the non-affine
bending regime. Here, we have shown that these growing boundary layers are
rather a consequence of a transition from an exponential to a linear force
profile in the boundary layers. This follows naturally from the fact that
non-affine deformations, $\delta_{\rm na}$, scale with the fiber length $l_f$
and not with the segment length $l_s$. As we have analyzed in detail in
Ref.~\cite{heu07condmat}, this scaling property, which originates in the network
architecture, can be understood within a ``floppy mode'' concept.  Therefore,
the growing boundary layers should be viewed as a consequence rather than the
driving force of the affine to non-affine transition.

\section{Force distribution}\label{sec:stress-distribution}

\begin{figure}[thb]
  \begin{center}
    \includegraphics[width=0.9\columnwidth]{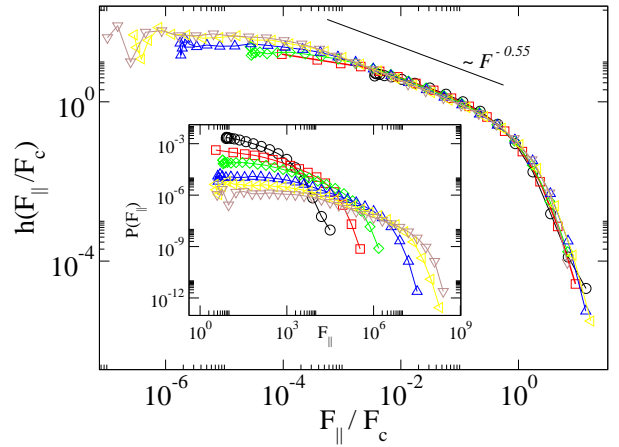}
\end{center}
\caption{Probability distribution $P(F_\parallel)$ (inset) and  
  scaling function $h$ in the non-affine bending regime with aspect
  ratio $r/l_f=5 \cdot10^{-6}$ for various densities $\rho
  l_f=20...100$. The force scale used to obtain the data collapse is
  $F_c=\kappa\rho_c^2(\delta\!\rho/\rho_c)^5$.}
   \label{fig:NAstressDistribution}
\end{figure}

We now turn to a discussion of the full probability distribution of axial
forces, instead of just the average value as we have done in the previous
section.  In Fig.~\ref{fig:NAstressDistribution} we display the distribution
function $P(F_\parallel)$ for various densities $\rho$ deep in the non-affine
regime.  Remarkably, the curves for different densities collapse on a single
master curve by using the scaling ansatz
\begin{equation}\label{eq:scaledForceDist}
   P(F_\parallel) = F_c^{-1} h(F_\parallel/F_c)\,,
\end{equation}
with the force scale $F_c = \kappa\rho_c^2(\delta\!\rho/\rho_c)^5$, where
$\delta\!\rho=\rho-\rho_c$.

Its appearance in Eq.~(\ref{eq:scaledForceDist}) indicates that $F_c$ is the
average \emph{axial} force. We now show that it is furthermore equivalent to the
average \emph{bending} force $F_c = \langle k_\perp(l_s)\cdot\delta_{\rm na}
\rangle$ that is needed to impose the non-affine bending deformation
$\delta_{\rm na}$ on a segment of bending stiffness $k_\perp$.

The average $\langle .\rangle$ is performed over the segment-length distribution
$P(l_s)$~\footnote{In the random network structure considered here, this
  distribution is exponential, $P(l_s)=\rho\exp(-l_s\rho)$.}, restricted to
those segments that are longer than the rigidity scale $l_{\rm min}$. Recall,
that $l_{\rm min} $ plays the role of the shortest deformed segment such that
shorter segments do not contribute to the averaging. The equivalence of both
expressions for the force scale becomes evident by writing the average
explicitly,
\begin{equation}\label{eq:avgForce}
  F_c = \int_{l_{\rm min}}^\infty\!dl_sP(l_s)k_\perp(l_s)\delta_{\rm na}\,,
\end{equation}
and inserting $k_\perp\sim l_s^{-3}$. The integral is dominated by its lower
limit, which leads to $F_c \sim \kappa (\rho l_f)^5/l_f^2$ as required.

% This force scale can be seen to be equivalent to the average force $F_c =
% \langle k_\perp(l_s)\cdot\delta_{\rm na} \rangle $ needed to impose the
% non-affine bending deformation $\delta_{\rm na}$ on a fiber segment.  The
% average $\langle .\rangle$ is performed over the segment-length distribution
% $P(l_s)$~\footnote{In the random network structure considered here, this
% distribution is exponential, $P(l_s)=\rho\exp(-l_s\rho)$.}, restricted to
% those segments that are longer than the rigidity scale $l_{\rm min}$. Recall,
% that $l_{\rm min} $ plays the role of the shortest deformed segment. The
% equivalence of both expressions for the force scale becomes evident by writing
% the average explicitly, 
% \begin{equation}\label{eq:avgForce}
%    F_c = \int_{l_{\rm min}}^\infty\!dl_sP(l_s)k_\perp(l_s)\delta_{\rm na}\,,
% \end{equation}
% and inserting $k_\perp\sim l_s^{-3}$. This leads to $F_c \sim \kappa (\rho
% l_f)^5/l_f^2$ as required. 

Hence, the force scale $F_c$ has been identified as the average force that
induces the non-affine bending of a fiber. At the same time it occurs in the
probability distribution of stretching forces as depicted in
Fig.~\ref{fig:NAstressDistribution}. This two-fold role reflects the interplay of
bending and stretching in the effective medium theory, where stretching forces
in one fiber have to be balanced by bending forces in its neighbors. We can thus
conclude that the effective medium picture is the appropriate description for
forces up to the threshold $F_c$, i.e. for the ``typical'' properties of the
system.

% Hence, the force scale $F_c$ has been identified as the average axial force that
% leads to the bending of fibers.  As an immediate consequence, it are precisely
% the forces up to this scale that enter the effective medium theory developed in
% the previous section.  In this picture, stretching forces in one fiber have to
% be balanced by bending forces in the surrounding medium to effectuate the
% observed force profiles.

Interestingly, for forces smaller than $F_c$ the probability distribution
displays an intermediate power-law tail $h(x)\sim x^{-\beta}$, with an exponent
$\beta\approx 0.55$, which as yet defies a simple explanation. For forces larger
than $F_c$, one may even speculate about the existence of a second power-law
regime with a much higher exponent $\beta'\approx5$.  While in this regime the
distribution does not seem to decrease exponentially, the available range of
forces is too small to reach any final conclusions as to the functional form.

\begin{figure}[thb]
  \begin{center}
    \includegraphics[width=0.9\columnwidth]{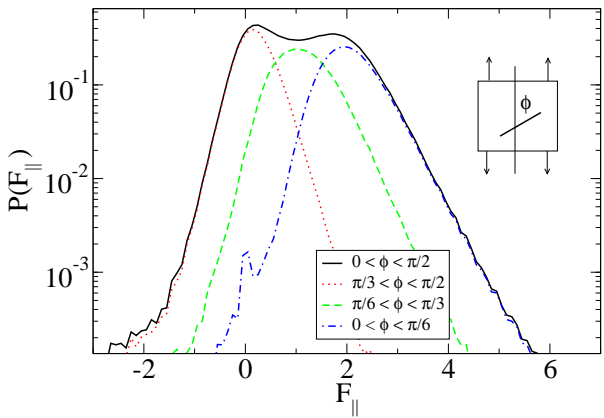}
\end{center}
\caption{Probability distribution $P(F_\parallel)$ of axial forces in  
  the affine stretching dominated regime for $\rho l_f=60$ and aspect
  ratio $r/ l_f= 0.05$.  The system is subject to uniaxial stretching,
  as indicated in the inset. Note the linear scale on the $x$-axis.
  The peak of the distribution is strongly correlated with the
  orientation, $\phi$, of the fiber relative to the principal
  stretching direction.}
   \label{fig:AstressDistribution}
\end{figure}

The axial forces in the affine stretching regime follow a completely
different probability distribution, as can be seen from
Fig.~\ref{fig:AstressDistribution}.  
The solid black line relates to the probability distribution of all segments,
irrespective of their orientation $\phi $, with respect to the imposed strain
field (in this case: uniaxial extension along the $y$-direction). The broken
lines only include segments with orientations $\phi$ in a particular interval,
as denoted in the figure legend. The tails of the distribution are exponential
(as has previously been observed in Ref.~\cite{ast94}), while the peak force
strongly depends on the orientation $\phi$ of the segments. The position of the
peak follows naturally from the assumption that segments undergo affine
deformations with $\delta_{\rm aff}=\gamma l_s\cos^2\phi$.  Since the resulting
affine forces $F_{\rm aff}(\phi)=k_\parallel(l_s)\delta_{\rm aff} \sim
\kappa\gamma\cos^2\phi$ do not depend on the length of the segments, one expects
only a single, orientation-dependent force, that is a delta-function
distribution $P(F,\phi)\sim \delta(F-F_{\rm aff}(\phi))$. The broadening of the
distribution relative to the singular delta-function can be rationalized with
the fact that the affine strain field can only fulfill force equilibrium if the
fibers are infinitely long. For finite fibers the force has to drop to zero at
the ends (as discussed above), leading to additional (non-affine) deformations
and therefore to a broadening of the peak.

\section{Force-chains}

In the preceding two sections we have characterized the occuring axial forces on
the scale $l_f$ of a single fiber as well as on the smaller scale of an
indivdual fiber segment $l_s$. We now proceed to discuss the properties of the
forces on a larger, mesoscopic scale.  To this end we probe the Green's function
of the system and impose a localized perturbation in the center of the network.
Fixing the boundaries, we displace a single crosslink and monitor the resulting
response of the network.  Similar studies have been conducted in
Ref.~\cite{hea05}, where ensemble averaging is used to discuss the applicability
of homogeneous elasticity theory.  Here, we focus on the individual network
realization to better characterize the spatial organization of the forces.

\begin{figure}[t]
  \begin{center}
    \includegraphics[width=0.85\columnwidth]{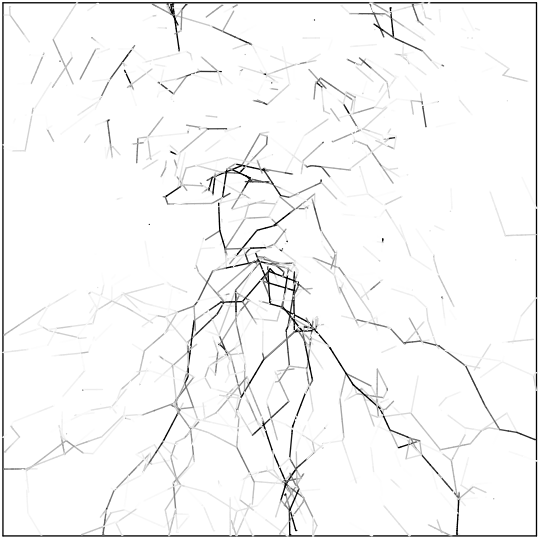}
\end{center}
\caption{Picture of a network in the non-affine bending dominated  
  regime. The grey-scale is chosen according to the axial force in the
  segment.  Force-chains are clearly visible, and follow a zig-zag
  course from the center towards the lower boundary.}
   \label{fig:forceChains}
\end{figure}

\begin{figure}[t]
  \begin{center}
    \includegraphics[width=0.85\columnwidth]{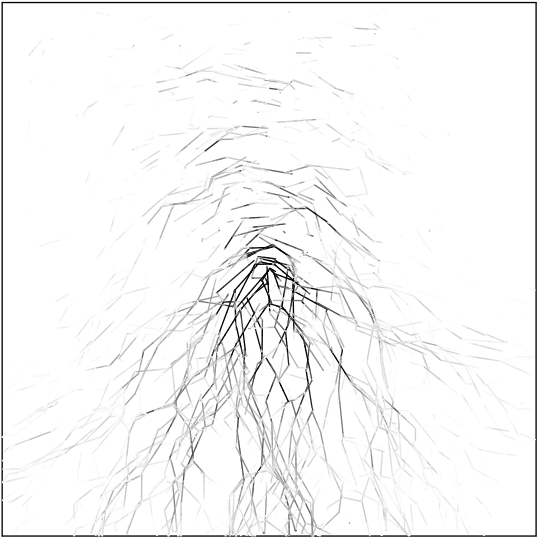}
\end{center}
\caption{Picture of a network in the affine stretching dominated regime. The
  grey-scale is chosen according to the axial force in the segment. No
  force-chains are present.}
   \label{fig:NOforceChains}
\end{figure}

In Figs.~\ref{fig:forceChains} and \ref{fig:NOforceChains} we show
pictures of a network, where the grey-scales of the segments are
chosen according to the magnitude of the forces they carry. The higher
the force the darker the segment. While the quenched random structure
is the same in both plots, the fiber aspect ratio $\alpha=r/l_f$ is
chosen such that the network lies deep in the non-affine bending
($\alpha = 10^{-5}$) or the affine stretching regime ($\alpha =
10^{-1}$), respectively. As is clearly visible in
Fig.~\ref{fig:forceChains} the non-affine response is characterized by
well defined paths of high forces that extend from the center, where
the force is applied, to the boundaries. These forces are transmitted
from fiber to fiber along a zig-zag course.  Upon comparison with the
distribution function for axial forces (like those shown in
Fig.~\ref{fig:NAstressDistribution}), we find that the force-chains
are constituted by forces with magnitude larger than $F_c$ (i.e. above
the intermediate power-law regime). In contrast, in the affine regime
a rather homogeneous distribution of forces is observed
(Fig.~\ref{fig:NOforceChains}).

\begin{figure}[t]
  \begin{center}
    \includegraphics[width=0.85\columnwidth]{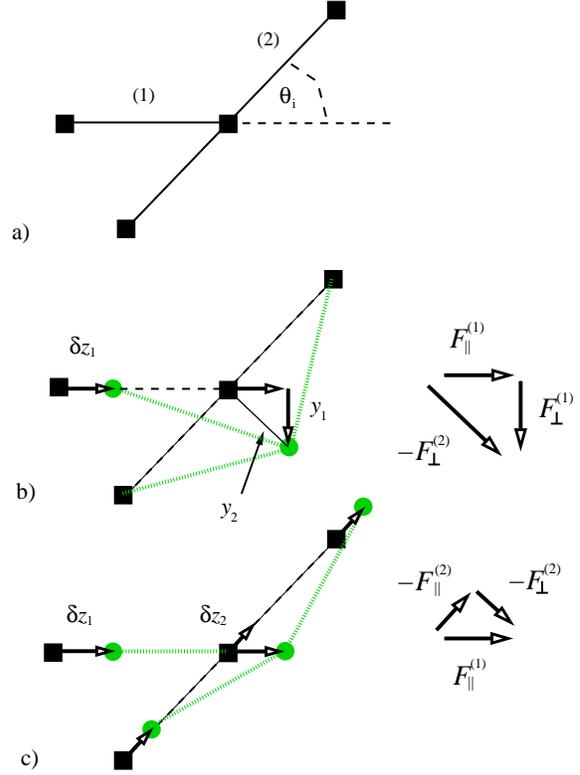}
\end{center}
\caption{Illustration of the two mechanisms of load transfer for 
  a fiber segment (1) ending at a crossing fiber (2).  The
  displacement $\delta\!z_1$ of the horizontal fiber leads to
  deformations in both fibers. The indirect load transfer, b), couples
  axial forces into bending deformations, while the direct transfer,
  c), transmits the axial force directly into axial forces in the
  neighboring fiber.}
   \label{fig:loadtransfer}
\end{figure}

The observation that the highest occuring forces are connected in chains,
suggests a mechanism that allows direct transmission of an axial load from one
fiber to the next. This has to be contrasted with the results of
Sec.~\ref{sec:cox-model}, where we have developed an effective medium picture
that constitutes an indirect load transfer, in which axial forces in one fiber
are balanced by transverse (bending) forces in the neighboring fiber. The
difference between both mechanisms becomes evident by considering the load
transfer at a fiber-fiber intersection that coincides with the end-point of one
of the two fibers; see Fig.~\ref{fig:loadtransfer} for an illustration.  An axial
load $F_\parallel^{(1)}$ that is coupled into fiber $(1)$ at its left end has to
be taken up by the crossing fiber $(2)$ at the distal end of fiber $(1)$.
Assume that this applied force in the horizontal fiber, $F_\parallel^{(1)}$, is
accompanied with an axial displacement $\delta\!z_1$, which translates the fiber
along its axis.

In the indirect load transfer the axial force $F_\parallel^{(1)}$ is balanced by
bending forces $F_\perp^{(1,2)}$ oriented perpendicular to the contour of both
fibers. The force balance reads
\begin{equation}\label{eq:fbalance_indirect}
   F_\parallel^{(1)} + F_\perp^{(1)} = - F_\perp^{(2)}\,.
\end{equation}
The angle between both fibers being $\theta$, the associated bending
displacements are (Fig.~\ref{fig:loadtransfer}b)
\begin{equation}\label{eq:y_perp}
  y_1 = -\delta\!z\cot\theta\,,\quad y_2 = \delta\!z/
  \sin\theta\,.
\end{equation}
%  $y_1 = -\delta\!z\cot\theta$ and $y_2 = \delta\!z/
% \sin\theta$
This mechanism forms the basis of the effective medium theory applied in
Sec.~\ref{sec:cox-model} to explain the average force profile along the fibers.
It has also been shown to allow a direct calculation of the macroscopic elastic
moduli of the network~\cite{heu06floppy,heu07condmat}.

We now proceed to discuss the second mechanism. It can immmediately be seen from
the $\theta$-dependence that the bending displacements $y_{1/2}$ of
Eq.(\ref{eq:y_perp}) can become increasingly large, if the fibers intersect at
small enough angles, $\theta\to0$. Then, the network responds in a different
way. As can be inferred from Fig.~\ref{fig:loadtransfer}c the large bending
contributions may be avoided by additionally moving the secondary fiber along
its own axial direction, $\delta\!z_2 = \delta\!z_1\cos\theta$. This leaves the
primary fiber straight while the bending displacement in the secondary fiber
becomes $y_2 = \delta\!z_1\sin \theta$.  The force balance must therefore be
written as
\begin{equation}\label{eq:fbalance_indirect}
   F_\parallel^{(1)} = - (F_\parallel^{(2)} + F_\perp^{(2)})\,.
\end{equation}
The smaller the angle $\theta$ the larger is the contribution of
$F_\parallel^{(2)}$, which is just the forward ``scattering'' of
forces, seen in Fig.~\ref{fig:forceChains}.

The existence of force-chains is intimately connected to the presence of long
fibers, $\rho l_f \gg 1$. Since the coordination can be written as
$z=z_c(1-O(\rho l_f)^{-1})$~\cite{heu07condmat}, the system will become
completely rigid by increasing the fiber length, $l_f \to \infty$.  There, the
force scale $F_c$ diverges and axial forces imposed on a fiber will propagate
along the fiber completely uncorrelated with forces on its neighbours.  As the
number of intersections $n_{\rm cl} \sim \rho l_f$ per fiber is proportional to
the fiber length, the probability distribution for the angle between two
intersecting fibers is sampled to an increasing degree. In particular, the
smallest angle that will occur in a finite sample of $n_{cl}$ intersections is
calculated from
\begin{equation}\label{eq:theta_min}
  \int_0^{\theta_{\min}}d\theta P(\theta) = 1/n_{cl}\,,
\end{equation}
as $\theta_{\min} \sim l_f^{-1/2}$, where we have used $P(\theta) \sim
\theta$ for small values of $\theta$. With increasing fiber length
ever smaller angles occur. As explained above, the presence of small
angles necessarily lead to forward scattering of axial forces, and
thus to the emergence of force-chains. The presence of force-chains is
therefore a consequence of the special geometry of the network that
allows the fibers to intersect at arbitrarily small angles.  To
support this view, we have conducted additional simulations in which
the localized perturbation is applied in varying directions. As a
result, the structure of the force-chains remain unchanged while their
amplitudes are modulated.

\section{Conclusion and Outlook}

In conclusion, we have characterized in detail the properties of forces occuring
in two-dimensional random fiber networks. We have shown that the previously
identified \emph{rigidity scale} $l_{\min}$, in addition to the \emph{structural
  scale} of the fiber length $l_f$, governs the occurence of stretching forces
in an elastic regime, where the energy derives from the bending mode only
($k_\parallel/k_\perp \to \infty$).  The probability distribution of forces
shows scaling behaviour with the force scale $F_c =
\kappa\rho_c^2(\delta\!\rho/\rho_c)^5$ that can be identified with the average
force needed to deform a fiber segment by the non-affine deformation
$\delta_{\rm na} \sim \gamma l_f$.

Two types of force transmission have been identified. Forces up to the scale
$F_c$ are transmitted from one fiber to the next by an \emph{indirect
  mechanism}, where stretching forces are balanced by bending forces and vice
versa. This is best illustrated by the action of a lever-arm that tries to
transmit its bending load to the neighboring fiber, which subsequently starts to
stretch (see $ {\cal K}_{\rm ext}$ in Fig.\ref{fig:illustration}). This
mechanism of force transmission can be used to understand the average force
profile along a fiber, and also forms the basis for the calculation of the
scaling properties of the elastic modulus~\cite{heu06floppy,heu07condmat}. In a
second \emph{direct mechanism} axial forces are also transmitted directly to
their neighbors, giving rise to highly localized force-chains that are
heterogeneously distributed throughout the sample. This mechanism is only active
for forces larger than $F_c$. In contrast to the indirect mechanism, which
probes the center of the force distribution and therefore typical properties of
the network, the direct mechanism reflects the properties of the extremal values
of the distribution.

The observation of force-chains suggests an analogy with granular
media~\cite{JaegerRMP96}. The scale separation $k_\parallel \gg k_\perp$ implies
that the fibers behave as if they were inextensible. The network of inextensible
segments may therefore loosely be viewed as the contact network of a system of
rigid grains.  Due to the low coordination number, the fiber network has to be
stabilised by the action of the bending mode directed perpendicularly to the
fiber axis.  This bears some similarity to a friction force in granular systems,
which is directed tangentially to the grain surfaces.  Indeed, it has been
argued that stable granular systems with a coordination as low as found here may
only be achieved if frictional forces are taken into account~\cite{unger05}. The
occurence of force chains in our ``frictional'' system is, however, due to the
vicinity of the isostatic point with regard to the \emph{frictionless} system
($z_c=4$). Since the coordination may be written as $z=4(1-O(\rho l_f)^{-1})$
the isostatic point is reached by increasing the fiber length $l_f\to\infty$. 
% It remains to be seen whether this loose analogy with granular media may be
% applied to other fibrous systems, e.g.  networks of semiflexible polymers.
% Possibly the results presented here may be used to better understand the
% properties of force propagation in granular media. They may also contribute to
% the debate whether it is at all possible to understand granular systems on the
% basis of elasticity theories.

%\begin{acknowledgments}
We gratefully acknowldege fruitful discussions with Klaus Kroy. Financial
support of the German Science Foundation (SFB486) and of the German Excellence
Initiative via the program "Nanosystems Initiative Munich (NIM)" is gratefully
acknowledged.
%\end{acknowledgments}

% %
% % For one-column wide figures use
% \begin{figure}
% % Use the relevant command for your figure-insertion program
% % to insert the figure file.
% % For example, with the option graphics use
% \resizebox{0.75\columnwidth}{!}{%
%   \includegraphics{leer.eps}
% }
% % If not, use
% %\vspace{5cm}       % Give the correct figure height in cm
% \caption{Please write your figure caption here}
% \label{fig:1}       % Give a unique label
% \end{figure}
% %
% % For two-column wide figures use
% \begin{figure*}
% % Use the relevant command for your figure-insertion program
% % to insert the figure file. See example above.
% % If not, use
% \vspace*{5cm}       % Give the correct figure height in cm
% \caption{Please write your figure caption here}
% \label{fig:2}       % Give a unique label
% \end{figure*}
% %
% % For tables use
% \begin{table}
% \caption{Please write your table caption here}
% \label{tab:1}       % Give a unique label
% % For LaTeX tables use
% \begin{tabular}{lll}
% \hline\noalign{\smallskip}
% first & second & third  \\
% \noalign{\smallskip}\hline\noalign{\smallskip}
% number & number & number \\
% number & number & number \\
% \noalign{\smallskip}\hline
% \end{tabular}
% % Or use
% \vspace*{5cm}  % with the correct table height
% \end{table}
%
% BibTeX users please use
% \bibliographystyle{}
% \bibliography{}

\end{document}